\documentclass[conference]{IEEEtran}
\IEEEoverridecommandlockouts
\usepackage{cite}
\usepackage{amsmath,amssymb,amsfonts}
\usepackage{algorithmic}
\usepackage{graphicx}
\usepackage{textcomp}
\usepackage{xcolor}
\def\BibTeX{{\rm B\kern-.05em{\sc i\kern-.025em b}\kern-.08em
    T\kern-.1667em\lower.7ex\hbox{E}\kern-.125emX}}

\usepackage{caption}
\usepackage{subcaption}
\begin{document}

\title{Dynamic Sizing of Frequency Control Ancillary Service Requirements for a Philippine Grid}

\author{\IEEEauthorblockN{Elgar John S. del Rosario$^{1}$, Jordan Rel C. Orillaza$^{2}$}
\IEEEauthorblockA{\textit{Electrical and Electronics Engineering Institute} \\
\textit{University of the Philippines Diliman}\\
Quezon City, Philippines \\
$^{1}$elgar.john.del.rosario@eee.upd.edu.ph, $^{2}$jordan.orillaza@eee.upd.edu.ph}
}

\maketitle

\begin{abstract}
Sizing frequency control ancillary service (FCAS) requirements is crucial for the reliable operation of power systems amid a continuous influx of variable renewable energy (VRE) generation. Reserve sizing is especially pertinent for the Philippine grids due to an expected transition to new FCAS classifications established by its Grid Code.
In lieu of the existing deterministic formulation, this work proposes a dynamic approach for sizing secondary and tertiary reserves that accounts for the stochasticity and variability of load demand and VRE.
We propose a method where historical power imbalances were calculated and clustered according to the time and day of week they occurred.
The conditional probabilities of forecast and noise errors were characterized using kernel density estimation.
Recursive convolution was performed to obtain the total reserve requirement probability distribution.
The method was tested on Visayas grid's historical system operation data and used target reliability levels on the error distributions to size upward and downward reserve needs.
Finally, the methodology was extended to demonstrate through a numerical experiment that sizing FCAS at temporal resolutions higher than one-hour, e.g., five-minute, provides the benefit of shrinking the required capacities by as much as 86.2\% compared to current deterministic FCAS sizing.
\end{abstract}


\section{Introduction}

Continuous and precise frequency control is critical in maintaining power system reliability. Frequency deviations are counteracted by deploying active power control reserves, also known as frequency control ancillary services (FCAS) \cite{NREL2011}. The adequate sizing of reserves is crucial as, on the one hand, undersizing can have serious negative impact, including load shedding, renewable energy curtailment, equipment damage, and in the worst scenario, blackouts. Reserve procurement costs, on the other hand, will rise if reserves are oversized. The Philippine electric power industry is also about to activate a reserve market, hence FCAS sizing is more critical \cite{DOEDC2021}. 

Owing to the increasing grid integration of variable renewable energy (VRE) sources, such as wind and solar PV generation, quantifying reserve requirements considering VREs has received considerable attention in the last decade \cite{Holttinen2012}. While VRE growth promotes the development of a clean and sustainable grid, VREs are also nonsynchronous, highly fluctuating and difficult to predict, and could induce higher levels of FCAS requirements \cite{Hirth2015}. Dynamic methods of sizing FCAS requirements due to the increasing integration of VREs using kernel density estimation \cite{Jost2015}, k-nearest neighbors and k-means clustering \cite{Bucksteeg2016, DeVos2019}, and dynamic Bayesian belief networks \cite{Fahiman2019} have been proposed in the literature. In contrast with deterministic approaches that are typically based on rules of thumb, these dynamic probabilistic approaches vary reserve requirements depending on expected system conditions, considering the severity and probabilities of a range of potential power imbalances.

In the Philippines, reserves are still sized based on deterministic rules. Presently the Philippine system operator sizes three types of FCAS on a day-ahead basis: regulating, contingency, and dispatchable reserves. Regulating reserves are set at 4\% of the hourly forecast demand in each hourly dispatch interval, which is based on intra-hour load variations in 2010 \cite{ASPP2011}. Meanwhile, contingency and dispatchable reserve levels are determined by the highest and second highest generating power outputs expected to be online, in accordance with the N-1 and N-1-1 criteria, respectively.

The current reserve classifications lack requirements for primary frequency response, often resulting in the activation of automatic load dropping (ALD) schemes as a first line of defense against large disturbances \cite{PGC2016}. To address this gap, a revision of the Philippine Grid Code of 2016 prescribes new FCAS classifications according to the hierarchy of frequency control. In the new FCAS classification, primary reserves must provide primary frequency response via governor control, secondary reserves must operate to restore the system frequency to 60 Hz either under automatic generation control (AGC) or manually upon the command of the system operator, and tertiary reserves aim to replenish depleted secondary reserves \cite{PGC2016}. This development in FCAS framework necessitates a new approach in sizing reserves.

 Furthermore, current practices have yet to determine the incremental reserve needs introduced by the growing VRE penetration. As a country indigenously rich in VRE resources, the Philippines targets to achieve 35\% renewable energy penetration by 2030 and 50\% by 2040 \cite{NREP}.
In NREL's Greening the Grid report, high renewable energy scenarios for the Luzon-Visayas system of the Philippines by 2030 were investigated and the issue of adequate reserve provision was raised \cite{Barrows}.
However, the said study retained an assumption of deterministic reserve rules that govern system operations until now due to the absence of detailed reserve requirements.

The increasingly stochastic nature of system operations demand dynamic, data-driven and probabilistic approaches that consider the additional variability and uncertainty introduced by VREs. Considering this fact, this work proposes a dynamic probabilistic approach as an appropriate methodology for sizing reserves for frequency restoration, particularly, secondary and tertiary reserves, in the context of the Philippines. The method is based on clustering imbalances according to the hour and day they happened and kernel density estimation. 
    
Additionally, a recent enhancement of the Philippine Wholesale Electricity Spot Market (WESM) shifts the energy markets in the Luzon and Visayas grids from one-hour to five-minute intervals \cite{WESMRules}. While reserve schedules continue to be based on hourly requirements, the Department of Energy tasks the system operator to conduct studies on the determination of required reserves vis-\`a-vis the 5-minute dispatch interval implementation \cite{DOEDC2021}. This work further extends the aforementioned dynamic reserve sizing methodology to perform numerical experiments for subhourly sizing.

In Section 2, the proposed methodology is discussed. Section 3 provides the results and discussion on scenario simulations done on real Philippine grid operations data. Section IV concludes with recommendations for further research.

\section{Proposed FCAS Sizing Methodology} \label{method}

\begin{figure}[t]
\centerline{\includegraphics[width=\columnwidth]{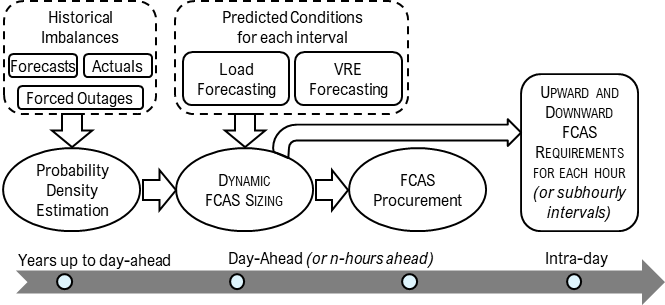}}
\caption{Proposed framework for dynamic FCAS sizing in the Philippines}
\label{framework}
\end{figure}

An overview of the proposed reserve sizing process is shown in Fig. \ref{framework}. Probability density estimation of various drivers for power imbalance is done based on historical values of power imbalances, which are mainly due to the discrepancy between the forecast and actual values of demand, wind and solar generation, as well as forced (unplanned) outages of power plants or interconnectors. The estimation process must be continuously done to incorporate new data and make the FCAS sizing process more accurate. On top of the estimated probability distribution functions (PDFs),  load and VRE forecasts are employed as basis for sizing. The status quo reserve sizing is done on the day-ahead for sizing reserves for each hour of the next day, but sizing and scheduling reserves closer to real time (\textit{n}-hours ahead) may prove to benefit forecast accuracy, especially when reserves are sized for subhourly intervals. The result of FCAS sizing shall then be the basis of FCAS procurement and the separate upward and downward capacities reserved for each intra-day interval.

\begin{figure}[t]
\centerline{\includegraphics[width=0.7\columnwidth]{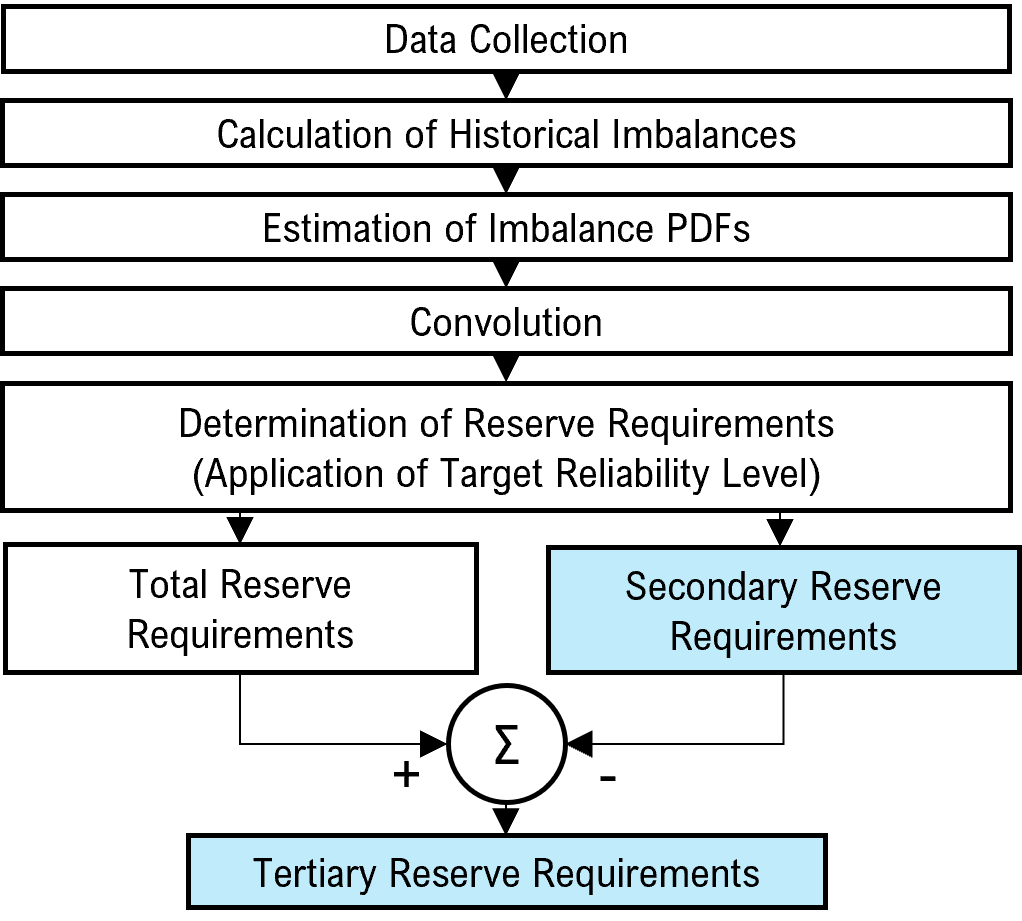}}
\caption{Overview of Proposed Methodology}
\label{overview}
\end{figure}

An overview of the methodology for reserve sizing is shown in Fig. \ref{overview}. The necessary data are collected for the calculation of historical imbalances. The historical imbalances are used to estimate the power imbalance PDFs. The convolution of these power imbalance PDFs produce PDFs for total and secondary reserve needs. The reserve requirements are determined by applying a target reliability level, referring to the acceptable percentage of time that there could be a shortage of reserves. The tertiary reserve requirement is taken as the difference between the derived total and secondary reserves. The following sections expound on these steps.

\subsection{Estimation of Error Probability Distributions}

\subsubsection{Forecast and Noise Errors}

\begin{figure}
    \centering
    \begin{subfigure}[b]{0.24\textwidth}
        \centerline{\includegraphics[width=\textwidth]{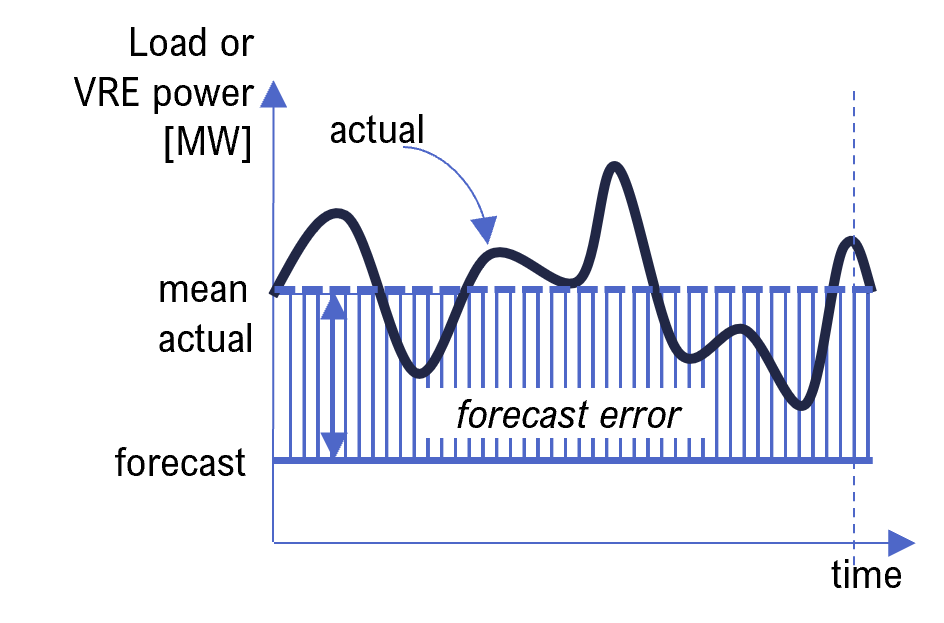}}
        \caption{Forecast Errors}
        \label{forecast_error}
    \end{subfigure}
\hfill
    \begin{subfigure}[b]{0.24\textwidth}
        \centerline{\includegraphics[width=\textwidth]{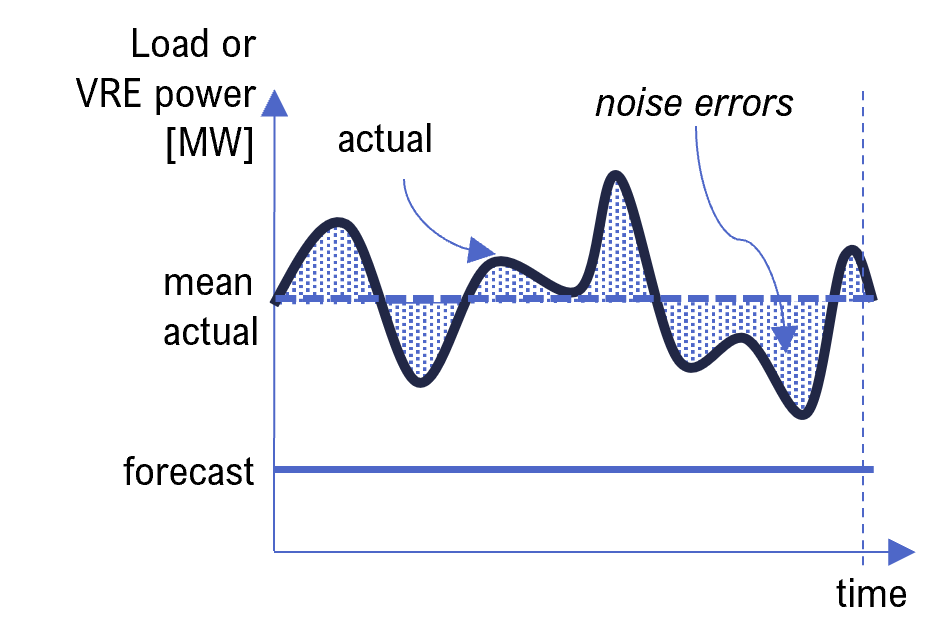}}
        \caption{Noise Errors}
        \label{noise_error}
        \end{subfigure}
    \caption{Calculation of Forecast and Noise Errors}
\end{figure}

Historical time series of load, wind and solar PV forecast and actual values are used to calculate forecast and noise errors. The load forecast error is the deviation of the mean value of load in a time interval from the forecast value, whereas the load noise error (or ``load oscillation") is the deviation of the actual value from the interval mean \cite{Maurer2009}. Calculating forecast and noise errors are depicted in Figs. \ref{forecast_error} and \ref{noise_error}.

To avoid making strong assumptions on the shape of the underlying probability distributions, this work uses kernel density estimation to estimate the PDFs of wind, solar and load forecast and noise errors as in \cite{Bucksteeg2016}. Kernel density estimation has been extensively used to characterize wind speed distributions and wind forecast errors in the literature \cite{ZhangWang}. Given the forecast and noise error values, $\epsilon_{i,t}$, the probability distribution function $f_i(u)$ is given by:

\begin{equation}
    f_i(u) = \frac{1}{T{h_i}} \sum_{t=1}^{T} K \left(\frac{u-\epsilon_{i,t}}{h_i}\right)
    \label{kde}
\end{equation}

where $K(\cdot)$ is the \textit{kernel smoothing function}; $u$, the evaluation point for function $f_i$;
$i \in [1, 24\times7]$ denotes an index to a cluster, which, in this work, is the hour and day of week of the error;
$t$, an index for time step; and
$T$, the maximum time step.
The kernel smoothing function is assumed to be a normal probability density function \cite{Jost2015, Bucksteeg2016}. The optimal bandwidth is proportional to the standard deviation of the cluster $i$, $\sigma_i$, and can be calculated from the empirical data as \cite{Bowman}:

\begin{equation}
h_i=\left(\frac{4}{3T}\right)^{0.2}\sigma_i.
\end{equation}

From kernel density estimation, we get six sets of PDFs, representing forecast and noise errors for load, wind and solar PV generation for each hour of the week: $f^x_y$ where $x \in \{\text{``forecast'', ``noise''}\}$ and $y \in \{\text{``load'', ``wind'', ``solar''}\}$.

\subsubsection{Power Plant Outages}
The forced outage rate, $FOR$, of a single generating plant or unit is calculated from the number of hours it \textit{was} on forced outage, $N_{outage}$, and the total number of hours in the period considered, $N_{hours}$:

\begin{equation}
FOR = \frac{N_{outage}}{N_{hours}}.
\end{equation}

The forced outage probability, $FOP$, reflects the probability of generation capacity \textit{going} out, instead of \textit{being} out \cite{Doherty2005}:
\begin{equation}
FOP =\frac{FOR}{MTTR}
\end{equation}
where $MTTR$ refers to the mean time to repair.
In determining the total capacity outage distribution, the FOP of each generating plant or unit is first calculated. 
The error distribution of a single plant is a piecewise-defined function

\begin{equation}
f_{j}^{\text{outage}}(X) =
    \begin{cases} 
    FOP_j, & \text{if } X= 0 \\
    1-FOP_j, & \text{if }X = P^\text{rated}_j
    \end{cases}
    \label{outageq}
\end{equation}
where $j$ is an index for a generating station and $P^\text{rated}_j$ is the generating capacity of station $j$. Then the error distributions of all power stations are recursively convolved to create a single error distribution\cite{Jost2015}, the total capacity outage distribution, $f^{\text{outage}}_\text{total}$:
\begin{equation}
f^\text{outage}_\text{total} = f^{\text{outage}}_{1} * f^{\text{outage}}_{2} * ... * f^{\text{outage}}_{N_{\text{gen}}-1} * f^{\text{outage}}_{N_{\text{gen}}}.
 \label{convoutages}
\end{equation}
where $N_{\text{gen}}$ is the total number of generating stations and $*$ is a shorthand operator for convolution. Convolution is defined for two random variables $\textbf{X}$ and $\textbf{Y}$ and their sum $\textbf{Z} = \textbf{X} + \textbf{Y}$, with their respective PDFs $f_X, f_Y, f_Z$:

\begin{subequations}
\begin{equation} 
f_Z = f_X * f_Y \label{convshort}
\end{equation}
\begin{equation} 
f_{Z}(z)=\sum _{k=-\infty }^{\infty }f_X(k) \cdot f_Y(z-k)
\end{equation}
\end{subequations}

\subsection{Determination of Reserve Requirements}

\subsubsection{Convolution of Error PDFs}
As each power imbalance driver is treated as a random variable, convolution is employed to determine the PDF of the sum of these imbalance drivers. The total reserve PDF is derived from convolution of the PDFs of all the defined power imbalance drivers: 
\begin{equation} 
\begin{split}
f^\text{total}_\text{reserve} = f^\text{forecast}_\text{load} * f^\text{noise}_\text{load} * f^\text{forecast}_\text{wind} * f^\text{noise}_\text{wind}\\
*f^\text{forecast}_\text{solar} * f^\text{noise}_\text{solar} *f^\text{outage}_\text{total} \label{totres}.
\end{split}
\end{equation}

\begin{figure}[t]
\centerline{\includegraphics[width=\columnwidth]{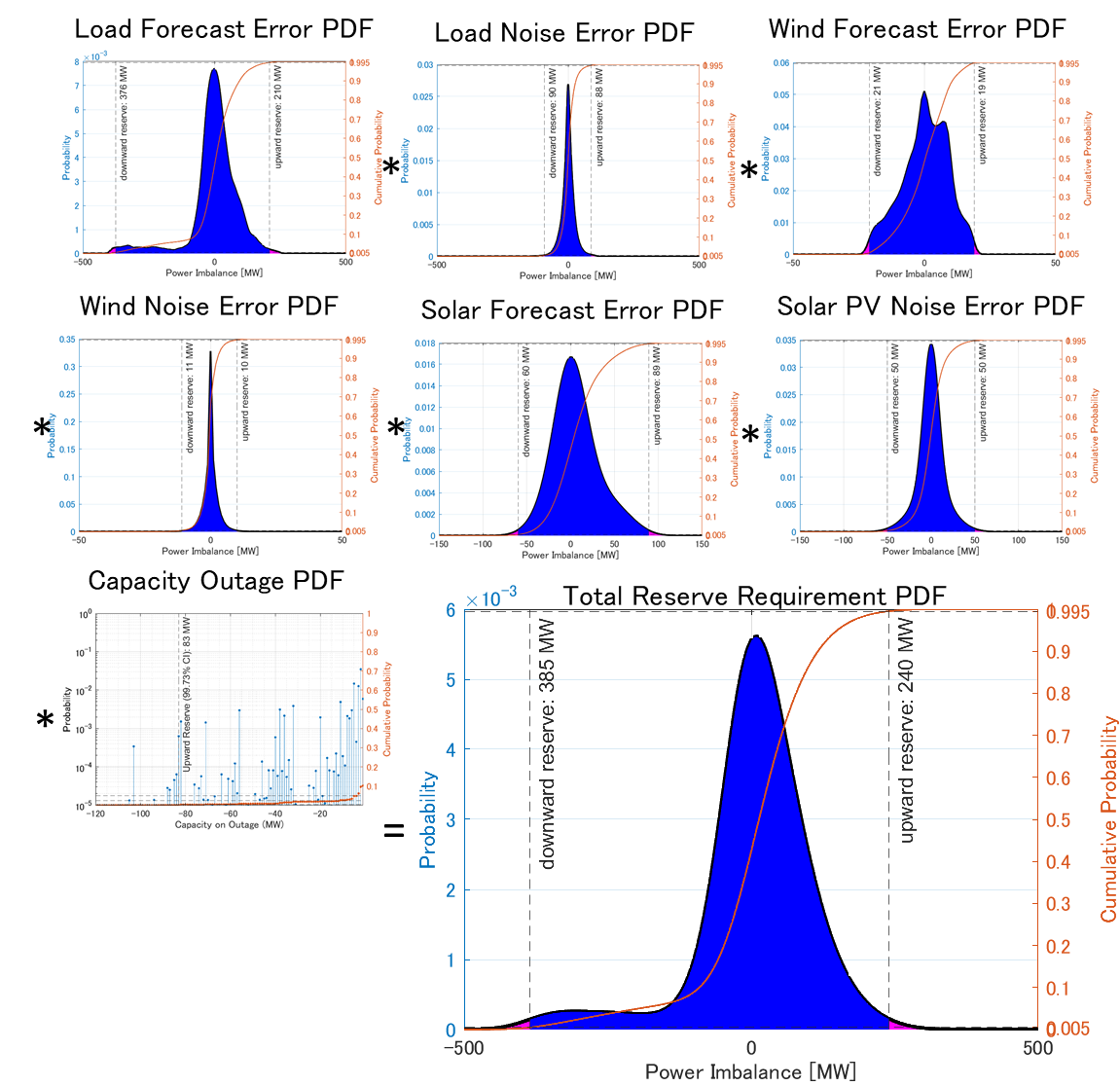}} 
\caption{Power imbalance driver PDFs convolved to derive the PDF of total reserve needs. Data based on 2018-2019 Visayas system operations. }
\label{subhourly}
\end{figure}

Figure 4 depicts the convolution of seven individual PDFs. Next, the secondary reserve PDF is the convolution of the PDFs of only the power imbalance drivers that need to be handled by secondary control, namely noise errors (fast fluctuations of demand and VRE) and capacity outages \cite{Maurer2009}:
\begin{equation}
\begin{split}
f^{\text{secondary}}_{\text{reserve}} = f^{\text{noise}}_{\text{load}} * f^\text{noise}_\text{wind} * f^\text{noise}_\text{solar} * f^\text{outage}_\text{total}.
\end{split}
\end{equation}

\subsubsection{Scaling of Error PDFs}

The predicted demand, wind and solar PV capacities for future scenarios are used to determine the reserve requirements.
The PDFs of the power imbalance drivers are scaled by a growth factor equivalent to the ratio of the forecast quantity, $P^\text{future}_i$ [MW] to its historical peak value $P^{\text{base}}_i$ [MW]. Furthermore, to account for future improvements in forecast performance, a forecast improvement factor, $k^\text{forecast}_i$, can be multiplied to the error values to get the scaled error values $\epsilon_{i}^\text{future}$.
        
\begin{equation} \label{Quantities for Future Scenarios}
\epsilon_{i}^\text{future} = \epsilon_{i}^\text{base} \cdot k_i^\text{forecast} \cdot \frac{P^\text{future}_i}{P^\text{base}_i}
\end{equation} \label{futures}
 
\subsubsection{Application of Reliability Margins}

The reliability margin $\rho_{margin}$, also referred to as the security level, is a predefined target that represents the fraction of time, typically over a one-year period, in which reserves are adequate to serve the demand \cite{Hirth2015, Bucksteeg2016}. The deficit ($\rho_\text{deficit}$) and surplus ($\rho_\text{surplus}$) probabilities correspond to the the fractions of time that shortage of downward and upward reserves, respectively, are acceptable \cite{Maurer2009}, and they are set to be equal in this work:
\begin{equation}
\rho_\text{deficit} = \rho_\text{surplus} = \frac{100\% - \rho_\text{margin}}{2}.
\label{rels}
\end{equation}

The cumulative distribution functions $F$ of the total and secondary reserve are derived as follows:
\begin{equation}
F_{\text{reserve}}(z)=\int_{-\infty}^{z}f_{\text{reserve}}(z) dz.
\end{equation}
The reserve requirements $R$ are determined by seeking the values of $F$ that satisfy the following inequality conditions with respect to the set values of $\rho_{\text{deficit}}$ and $\rho_{\text{surplus}}$:

\begin{subequations}
\begin{equation}
    F_{\text{reserve}}\left(R_{\text{up}}\right)\leq 1-\rho_{\text{surplus}} \label{one}
\end{equation}
\begin{equation}
    F_{\text{reserve}}\left(R_{\text{down}}\right) \geq \rho_{\text{deficit}} \label{two}
\end{equation}
\end{subequations}
Equations \ref{one} and \ref{two} apply to both total and secondary reserve requirements.
Finally, tertiary reserves are taken as the difference between the total and secondary reserve requirements:
\begin{equation}
R^{\text{tertiary}} = R^{\text{total}} - R^{\text{secondary}}. \label{three}
\end{equation}
Equation \ref{three} applies to both upward ($R_{\text{up}}$) and downward ($R_{\text{down}}$) reserve requirements. The proposed separation scheme between secondary (fast-response) and tertiary (slow-response) reserves is considered cost-effective because it reduces secondary reserves, which are generally more expensive than tertiary reserves \cite{DeVos2013}.

\section{Results and Discussion} \label{resultsch}

The Philippines is comprised of three synchronous grids, Luzon, Visayas, and Mindanao, each with a regional system operator with the National Grid Corporation of the Philippines (NGCP).
Reserve requirements are determined for each region separately at present.
This section demonstrates the reserve sizing using two years' worth of data for the Visayas grid, which has the highest VRE share among the three grids.
In 2021, solar accounted for 12.5\% share of total capacity and 4.01\% share of total gross generation, whereas wind accounted for a 2.4\% share of total capacity and 1.07\% share of total gross generation in the Visayas grid \cite{DOEStats}.

A one-week representative scenario is developed from real Visayas system operations data.
The forecast values for each hour of the week are set to be equivalent to the maximum historical values of forecast demand and wind and solar PV power generation that were observed in the same hour of the week in years 2018-2019.
$k_i^\text{forecast} = 1$ is assumed, which implies that the error distributions did not change from the historical period used.
The following historical data from the NGCP for the Visayas grid were used:
\begin{enumerate}
\item hourly forecast and per-minute actual system demand from 1 January 2018 to 31 December 2019.
\item hourly forecast VRE generation from 2 wind and 11 solar PV plants from 1 Jan. 2018 to 1 Dec. 2019.
\item per-minute actual VRE generation from 1 Jan. 2018 to 31 Dec. 2019.
\item forced outages from 1 Jan. 2016 to 31 Dec. 2019.
\end{enumerate}

\subsection{Static and Dynamic Secondary Reserve vs. Status Quo RR Requirements} \label{Dyn}

In this section, we compare the secondary reserve requirement with status quo regulating reserve requirement and the results of a static method. Secondary reserves and regulating reserves (RR) as currently analogous in Philippine practice as they are both intended to be operated under automatic generation control to restore the frequency to its nominal value. The RR requirement is set at 2\% of the forecast demand for upward regulation and 2\% of the forecast demand for downward regulation \cite{WESMCentral}.

On the other hand, static and dynamic methods require the same types of error distributions, but differ in the forecast data used to estimate the PDF and size the reserves. The static method is based on the peak forecast in the whole period for which reserves will be sized, and all historical values of an imbalance driver are included in the PDF estimation, regardless of its time of occurrence. The dynamic method is based on hourly forecasts. Every hour of the week has a distinct PDF for each imbalance driver, wherein only historical values that occurred in a similar hour of the week are included.

\begin{figure}[htbp]
\centerline{\includegraphics[width=\columnwidth]{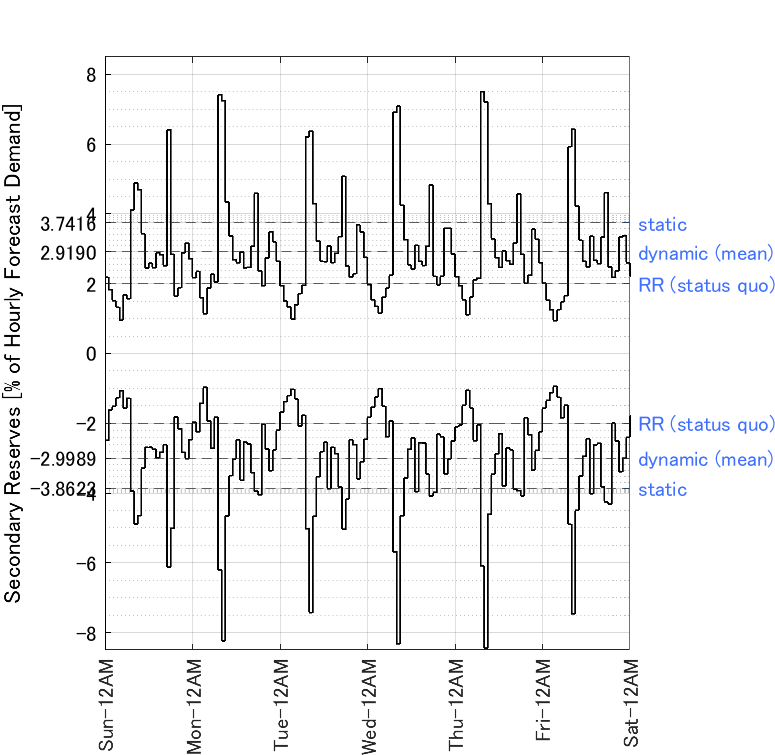}}
\caption{Dynamic Upward (+) and Downward (-) Secondary Reserves vs. Static and Regulating Reserve Requirements}
\label{secperc}
\end{figure}

The results of dynamic secondary reserve sizing is shown in Figure \ref{secperc}. The mean dynamic reserve requirements are lower than the static reserve requirements, as likewise observed in previous studies \cite{Jost2015, Bucksteeg2016}.
Furthermore, both static and mean dynamic reserve requirements are larger than the current $\pm$2\% requirement for regulating reserves.
Throughout the one-week scenario, it appears that most hours require levels higher than the current regulating reserve requirements.
Hours 1-7 and 20 provide exceptions as they are well within the status quo requirement on most days. Based on a 99.0\% reliability margin for a one-week Visayas grid scenario, the study reveals a need for asymmetric upward and downward reserves, which at some hours may be greater than the levels determined by existing practice. 

\subsection{Subhourly Reserve Sizing} \label{Subhr}
To contribute to the discussions on reserve requirements in view of the subhourly dispatch regime, we apply the proposed methodology to quantify the potential effect of varying the reserve lengths from one-hour to higher resolutions of 30-minute, 15-minute, and 5-minute.

As subhourly forecasts were not yet available in 2018-2019, we synthesized subhourly forecasts of demand and VRE generation using historical actual measurements. At the end of each dispatch interval, we assumed that subhourly forecast values are equivalent to the actual measurements.
Thus, only the power imbalances within the interval are allocated reserves.

Table \ref{subhour} shows the results of subhourly sizing for total reserves. Compared to hourly reserves, more granular reserves are considerably reduced. In particular, five-minute reserve sizing can reduce the total reserve requirements by as much as 86.2\% on average. The results suggest economic benefits from the decreased reserve need while still upholding the required level of system reliability.
More research is needed to determine the necessary adjustments in operations, as well as regulations, to ensure that dynamic reserve sizing of five-minute lengths, or subhourly lengths, is indeed feasible.

\begin{table}[htbp]
\caption{Average Dynamic Total Reserve Requirements at Subhourly Intervals and 99.0\% Reliability}
\begin{center}
\begin{tabular}{|c|c|c|c|c|}
\hline
\textbf{Time}&\multicolumn{2}{|c|}{\textbf{Downward}}&\multicolumn{2}{|c|}{\textbf{Upward}}\\
\cline{2-5}
\textbf{Res.} & \textbf{\textit{Mean[MW]}}& \textbf{\textit{\%Reduction$^{\mathrm{1}}$}}& \textbf{\textit{Mean[MW]}} &\textbf{\textit{\%Reduction$^{\mathrm{1}}$}}\\
\hline
60-min & 356.0 & 0\%  & 205.1 &       0\%   \\
30-min & 201.2 & -43.5\% & 122.3 & -40.4\% \\
15-min & 115.6  & -67.5\% & 75.3 & -63.3\% \\
5-min  & 49.1  & -86.2\% & 35.4  & -82.7\% \\ 
\hline
\multicolumn{5}{l}{$^{\mathrm{1}}$ =(mean subhourly total - mean hourly total)/(mean hourly total).}
\end{tabular}
\label{subhour}
\end{center}
\end{table}

\section{Conclusions and Recommendations} \label{concl}

A dynamic reserve sizing methodology is proposed as the Philippine grids transition to a new set of reserve classification and face increasing levels of VRE penetration. In comparison with the current allocation method, the proposed method indicate a larger capacities for reserves to achieve a 99.0 percent reliability. As the reserve requirements are anchored to a predetermined reliability target, further study is required to determine the optimal reliability level to aspire to, ideally considering the prices of reservation and activation.

Numerical experiments also revealed that dynamic five-minute reserve sizing can reduce reserve requirements by as much as 86.2\% compared to hourly requirements. New five-minute forecasts from recent operations and higher resolution actual data would be preferred in order to assess subhourly sizing more precisely.
Validating the effectiveness of the proposed method through pilot implementation and the evaluation of its effect on frequency quality can also be put forward.

\end{document}